\documentclass[prb,preprint]{revtex4-1} 

\usepackage{amsmath}  
\usepackage{amsfonts} 
\usepackage{graphicx} 
\usepackage{braket}
\usepackage{listings}
\usepackage{float}

\begin{document}


\title{Another formula for calculating Clebsch Gordan coefficients}


\author{Everardo Rivera-Oliva}
\email{everardo.rivera@cinvestav.mx}
\affiliation{Departamento de Física,\\
Centro de Investigación y de Estudios Avanzados del Instituto Politécnico Nacional,\\
P.O. box 14-740, C.P. 07000, Ciudad de México, México}


\date{\today}

\begin{abstract}
This article presents the derivation of a comprehensive formula for the Clebsch-Gordan coefficients in a quantum system. The formula is derived by employing the iterative application of angular momentum ladder operators on each defined angular momentum subspace to reconstruct the states in the total angular momentum base. The novelty aspect of this approach lies in the utilization of the $J{+}$ raising operator to reconstruct subspaces characterized by non-maximal total angular momentum, in contrast to the conventional Gram-Schmidt procedure typically employed in the standard literature.This enables us to derive a new formula that provides an alternative approach for computing these coefficients, complementing the existing ones found in the literature.
\end{abstract}

\maketitle 

\section{Introduction} 
The addition of angular momentum within the framework of Quantum Mechanics presents a persistent challenge for physicists. It begins with the conceptualization of the Hilbert space of a system, representing the sum of two angular momenta, as the direct product of the Hilbert spaces of the individual component systems. This process extends to identifying a Complete Set of Commuting Observables (C.S.C.O.) that labels the states in the compound system's Hilbert space and determines their associated spectrum. A significant difficulty lies in the transformation of basis in the compound Hilbert space, necessitated by the introduction of the Clebsch-Gordan coefficients.\\
The Clebsch-Gordan coefficients are the coefficients that emerge in the transformation of the tensor product basis of individual angular momenta $\ket{j_1,j_2,m_1,m_2}$ into the basis of the summation of angular momenta $\ket{J,M}$. The computation of these coefficients is intricate, and due to the absence of general formulas in contemporary Quantum Mechanics literature, many physicists remain unaware of their existence for calculative purposes.Racah\cite{Racah1942},Wigner\cite{wigner2012group},Schwinger\cite{schwinger1960angular} and others\cite{Sharp1960,ohara2001clebschgordancoefficientsbinomialdistribution,pain2024clebschgordancoefficientshypergeometricfunctions} have derived formulas for the Clebsch-Gordan coefficients through various methodologies. Although these expressions remain lengthy and intricate, they offer an advantage over the conventional methods presented in standard textbooks.
\\This study introduces an alternative approach to derive a general formula for Clebsch-Gordan coefficients, utilizing standard techniques from elementary Quantum Mechanics. The paper is structured as follows: Section \ref{s1} provides an overview of the general theory concerning the addition of angular momentum and presents the existing general formulas documented in the literature. In Section \ref{s2}, we perform the calculations to derive an alternative expression for the Clebsch-Gordan coefficients by iteratively applying angular momentum ladder operators within subspaces of fixed angular momentum. Finally, Section \ref{s3} delineates the conclusions and proposes future research directions.

\section{Overview of general theory of angular momentum}\label{s1}
The general theory of angular momentum has had several developers throughout history, but the core foundations are attributed to be done by Dirac \cite{dirac1981principles} and von Neuman \cite{von1955mathematical}, we'll be using a modern notation provided in any standard quantum mechanics book (see Cohen \cite{cohen2019quantum} for a more detailed discussion). \\
Consider a quantum system characterized by a state space $ \mathcal{E}$ and an angular momentum operator $J$ associated with the system. As established in Cohen\cite{cohen2019quantum} and others, it is feasible to construct a standard basis for $\mathcal{E}$ $\{\ket{k,j,m}\}$ consisting of common eigenvectors of $J^2,J_z$:
\begin{equation}\label{eq2.1}
\begin{split}
    &J^2\ket{k,j,m} = \hbar^2 j(j+1)\ket{k,j,m},\\
    &J_z\ket{k,j,m} = \hbar m\ket{k,j,m}.
\end{split}
\end{equation}
In this context, $k$ represents the eigenvalues of a collection of operators, which, together with $J^2,J_z$, constitute a Complete Set of Commuting Observables (C.S.C.O.). Additionally, we define the raising and lowering angular momentum operators indicated by $J_{+},J_{-}$ to be such that:
\begin{equation}\label{eq2.2}
            J_{\pm}\ket{k,j,m}=\hbar \sqrt{j(j+1)-m(m \pm 1)}\ket{k,j,m\pm1}.
\end{equation}
The collection of eigenvectors associated with a fixed $k,j$ delineates a $(2j+1)$ dimensional vector space, referred to as $\mathcal{E}(k,j)$. 

Consider a quantum system comprising the union of two subsystems, which we shall denote by indices 1 and 2. The corresponding state spaces $\mathcal{E}_1,\mathcal{E}_2$ possess standard bases $\{\ket{k_1,j_1,m_1} \}$ and $\{\ket{k_2,j_2,m_2} \}$, characterized by the common eigenvectors $J_1^2,J_{1z}$ and $J_2^2,J_{2z}$, respectively:
\begin{equation}\label{eq2.4}
    \begin{split}
        &J_1^2\ket{k_1,j_1,m_1}=\hbar^2 j_1(j_1+1)\ket{k_1,j_1,m_1},\\
        &J_{1z}\ket{k_1,j_1,m_1}=\hbar m_1 \ket{k_1,j_1,m_1},\\
        &J_{1\pm}\ket{k_1,j_1,m_1}=\hbar\sqrt{j_1(j_1+1)-m_1(m_1\pm1)}\ket{k_1,j_1,m_1\pm1}.
    \end{split}
\end{equation}

\begin{equation}\label{eq2.5}
    \begin{split}
        &J_2^2\ket{k_2,j_2,m_2}=\hbar^2 j_2(j_2+1)\ket{k_2,j_2,m_2},\\
        &J_{2z}\ket{k_2,j_2,m_2}=\hbar m_2 \ket{k_2,j_2,m_2},\\
        &J_{2\pm}\ket{k_2,j_2,m_2}=\hbar\sqrt{j_2(j_2+1)-m_2(m_2\pm1)}\ket{k_2,j_2,m_2\pm1}.
    \end{split}
\end{equation}
The state space of the entire system $\mathcal{E}$ is defined as the tensor product of each individual state space:
\begin{equation}\label{eq2.6}
    \mathcal{E}=\mathcal{E}_1\otimes\mathcal{E}_2.
\end{equation}
Subsequently, the basis for the entire state space is similarly represented by the tensor product of the corresponding subspaces:
\begin{equation}\label{eq2.7}
    \ket{k_1,k_2,j_1,j_2,m_1,m_2}=\ket{k_1,j_1,m_1}\otimes \ket{k_2,j_2,m_2}.
\end{equation}

The total angular momentum operator can be expressed in terms of the individual angular momentum operators associated with each subspace:
\begin{equation}\label{eq2.10}
\begin{split}
    &J=J_1+J_2,\\
    &J_z=J_{1z}+J_{2z}.
\end{split}
\end{equation}
Utilizing the basis constructed in Eq.\eqref{eq2.7}, a state $\ket{k_1,k_2,j_1,j_2,m_1,m_2}$ acts as a common eigenstate of $J_1^2,J_2^2,J_{1z},J_{2z}$, which is particularly effective for analyzing the individual angular momentum of each subsystem. However, it proves less advantageous for examining the total angular momentum $J$. Nevertheless, we can construct a common basis of eigenvectors of $J_1^2,J_2^2,J^2,J_z$. This newly formed basis differs from the previous one, given that $J^2,J_{1z},J_{2z}$ do not commute, thereby presenting the challenge of a basis transformation.
Since $J$ is an angular momentum operator that preserves the invariance of the subspace $\mathcal{E}(j_1,j_2)$, it follows that this subspace can be expressed as a direct sum of orthogonal subspaces characterized by specified $k,J$:
\begin{equation}\label{eq2.13}
    \mathcal{E}(j_1,j_2)=\sum_{\bigoplus}\mathcal(k,J).
\end{equation}
The common eigenvectors, denoted by $\{\ket{J,M}\}$, of $J^2,J_z$ with fixed eigenvalues $j_1,j_2$ are expressed as follows:
\begin{equation}\label{eq2.14}
    \begin{split}
        &J^2\ket{J,M}=\hbar^2 J(J+1)\ket{J,M}, \\
       & J_z\ket{J,M}=\hbar M\ket{J,M}.
    \end{split}
\end{equation}
where:
\begin{equation}\label{eq2.15}
|j_1-j_2|\leq J\leq j_1+j_2, \quad -(j_1+j_2)\leq M\leq j_1+j_2.
\end{equation}
Due to Eq.\eqref{eq2.13}, within every subspace $\mathcal{E}(j_1,j_2)$, the eigenvectors of $J^2,J_z$, as specified by $\ket{J,M}$, are expressed as linear combinations of the initial basis $\{\ket{j_1,j_2,m_1,m_2} \}$:
\begin{equation}\label{eq2.16}
    \ket{J,M}=\sum_{m_1=-j_1}^{j_1}\sum_{m_2=-j_2}^{j_2} C(j_1,j_2,m_1,m_2,J,M) \ket{j_1,j_2,m_1,m_2}.
\end{equation}
The coefficients $C(j_1,j_2,m_1,m_2,J,M)$ involved in the linear combination are referred to as the Clebsch-Gordan coefficients. Upon determination of these coefficients, the transition from the basis of individual angular momentum to that of the combined angular momentum is achieved.

Various methodologies for the computation of Clebsch-Gordan coefficients have been established historically. Wigner \cite{wigner2012group} derived an expression for representing the Clebsch-Gordan coefficients through the utilization of the Wigner 3j-symbols, as articulated by:
\begin{equation}\label{w1}
    C(j_1,j_2,m_1,m_2,J,M)=(-1)^{M+j_1-j_2}\sqrt{2J+1}\begin{pmatrix}
j_1 & j_2 & J\\
m_1 & m_2 & -M
\end{pmatrix}
\end{equation}
where the Wigner 3j-symbol can be expressed as:
\begin{equation}
    \begin{pmatrix}
j_1 & j_2 & J\\
m_1 & m_2 & -M
\end{pmatrix} = \frac{1}{\sqrt{(J+1)!}}\sum_{n}(-1)^n\prod_{i=1}^3\frac{\sqrt{(j_i+m_i)!(j_i-m_i)!(J-2j_i)!}}{n_i!(J-2j_i-n_i)!}
\end{equation}
where $n=n_1+n_2+n_3$, and the summation extends over all $n_i$ fulfilling $J-2j_1\geq n_i \geq0$. Schwinger\cite{schwinger1960angular} independently derived the same formula as Wigner, presented in Eq.\eqref{w1}, employing a distinct methodology that involved reformulating the angular momentum commutation relations in terms of the harmonic oscillator framework.\\
Racah\cite{Racah1942} developed a comprehensive formula, currently referred to as Racah's formula (for a contemporary derivation, see Bohm\cite{Bohm1986} or Edmonds\cite{edmonds1996angular}), to express the Clebsch-Gordan coefficients algebraically by employing the recursion relations that these coefficients adhere to. By adopting this methodology, he derived the following expression:
\begin{widetext}
\begin{equation}\label{racah1}
    \begin{split}
        C=&\delta_{m_1+m_2,M}\sqrt{\frac{(2j_1+1)(j_1+j_2-J)!(J+j_1-j_2)!(J+j_2-j_1)!}{(j_1+j_2+J+1)!}} \times\\
        &\sum_z \frac{(-1)^z\sqrt{(j_1+m_1)!(j_1-m_1)!(j_2+m_2)!(j_2-m_2)!(J+M)!(J-M)!}}{z!(j_1+j_2-J-z)!(j_1-m_1-z)!(j_2+m_2-z)!(J-j_2+m_1+z)!(J-j_1-m_2+z)!}
    \end{split}
\end{equation}
\end{widetext}
where the conditions for the summation index $z$ ensure that all factorial terms remain non-negative. Sharp\cite{Sharp1960} re-formulated Racah's equation by identifying the scalar eigenfunctions of the angular momentum operator, which were applied to eigenfunctions of complex variables. Furthermore, the works of O'hara\cite{ohara2001clebschgordancoefficientsbinomialdistribution} and Pain\cite{pain2024clebschgordancoefficientshypergeometricfunctions} have found ways to express specific Clebsch-Gordan coefficients in terms of properties of the binomial probability distribution and hypergeometric functions. In addition, some alternative formulas can be found in Wolfram\cite{wolfram-clebsch}.

\section{General formula for Clebsch-Gordan coefficients}\label{s2}
The approach we will employ to derive a comprehensive formula for Clebsch-Gordan coefficients entails the iterative application of $J_{\pm}$ operators on each subspace $\mathcal{E}(j_1,j_2)$ to reconstruct the states $\{\ket{J,M}\}$, by leveraging the characteristics of ladder operators acting within a specific subspace. The novelty of the approach will be encountered in the alternative construction of non-maximal angular momentum subspaces through the application of $J_{+}$ operator instead of following standard Gram-Schmidt procedure. Given the complexity of the expression, a ket in the $\{\ket{J,M}\}$ basis will be referred to as:
\begin{equation}\label{eq2.17}
    \ket{J,M}=\ket{j,m}_{J}.
\end{equation}
Moreover, given that the $J_{\pm}$ do not modify the total angular momentum eigenvalue, we will utilize the subsequent notation to represent the states within the tensor product of individual angular momentum associated with a particular subspace $\mathcal{E}(j_1,j_2)$:
 \begin{equation} \label{eq2.18}
\ket{j_1,j_2,m_1,m_2}=\ket{m_1,m_2}.
\end{equation}
The exhaustive notation $\ket{j_1,j_2,m_1,m_2}$ ought to be utilized only with reference to states that constitute components of separate subspaces $\mathcal{E}(j_1,j_2)$.
\subsection{Building the states with $J=j_1+j_2$}
We shall start by reconstructing all the states $\{\ket{J,M}\}$ characterized by total angular momentum $J=j_1+j_2$. The construction of the state possessing the maximum possible angular momentum, along with the highest z component of angular momentum, is determined by the following designation:
\begin{equation}\label{d7}
    \ket{j_1+j_2,j_1+j_2}_J=\ket{j_1,j_2}.
\end{equation}
By applying ${J}_{-}$ to Eq.\eqref{d7}, we derive the states within subspace $J=j_1+j_2$. Let us develop the general formulas pertaining to the application of ${J}_{-}$ upon an single system angular momentum state $\ket{j,m=j}=\ket{j,j}$:
\begin{equation}\label{d8}
\begin{split}
    &{J}_{-}\ket{j,m=j} = \hbar \sqrt{j(j+1)-j(j-1)}\ket{j,m=j-1},\\
    &{J}_{-}^2\ket{j,m=j}=\hbar^{2}\sqrt{j(j+1)-j(j-1)}\sqrt{j(j+1)-(j-1)(j-2)}\ket{j,m=j-2},\\
    &\vdots\\
    &{J}_{-}^{n}\ket{j,m=j} = \hbar^{n}\sqrt{\prod_{w=1}^{n}j(j+1)-(j-w+1)(j-w)}\ket{j,m=j-n}.
    \end{split}
\end{equation}
The product presented in Eq.\eqref{d8} can be simplified as follows:
\begin{equation}\label{d9}
    \begin{split}
    \prod_{w=1}^{n}j(j+1)-(j-w)(j-w+1) &=\prod_{w=1}^{n} j^2+j-(j^2-wj+j-wj+w^2-w) \\
    &=\prod_{w=1}^{n}j^2+j-j^2-j+2wj-w^2+w \\
    &=\prod_{w=1}^{n}2wj-w^2+w=\prod_{w=1}^{n}w(2j-w+1)\\
    &=\prod_{w=1}^n w \prod_{w=1}^{n}2j-w+1\\
    &=n!(2j-n-1)!=(n!)^2 \binom{2j}{n}.
    \end{split}
\end{equation}
Accordingly, Eq.\eqref{d8} can be reformulated as follows:\begin{equation}\label{d10}
    {J}_{-}^{n}\ket{j,j}_J=\hbar^n n!\sqrt{\binom{2j}{n}}\ket{j,j-n}_J.
\end{equation}
Utilizing Eq.\eqref{d9}, we obtain the expression related to the application of $J^n_{-}$ within $\ket{J=j_1+j_2,M=j_1+j_2}_J$, as delineated by:
\begin{equation}\label{d11}
    {J}_{-}^{n}\ket{J=j_1+j_2,M=j_1+j_2}_J=\hbar^{n}n!\sqrt{\binom{2j_1+2j_2}{n}}\ket{J=j_1+j_2,M=j_1+j_2-n}_J.
\end{equation}
But:
\begin{equation}\label{d12}
    {J}_{-}^{n}\ket{J=j_1+j_2,M=j_1+j_2}_J={J}_{-}^{n}\ket{m_1=j_1,m_2=j_2}.
\end{equation}
where in the tensor product base $J_{-}$ is expressed as follows:
\begin{equation}\label{d13}
    {J}_{-}={J}_{-1}+{J}_{-2}.
\end{equation}
${J}_{-1}$ operates within the Hilbert space $\mathcal{E}_1$, whereas ${J}_{-2}$ acts within the Hilbert space $\mathcal{E}_2$. Consequently, these operators commute:
\begin{equation}\label{d14}
    [{J}_{-1},{J}_{-2}]=0.
\end{equation}
Due to the commutation property, $J_{-}^n$ can be expressed as a binomial series:
\begin{equation}\label{d15}
    {J}_{-}^{n}=\left({J}_{-1}+{J}_{-2} \right)^n=\sum_{k=0}^{n}\binom{n}{k}{J}_{-2}^{n-k}{J}_{-1}^{k}.
\end{equation}
subsequently, the right-hand side of Eq.\ref{d12} can be articulated as follows:
\begin{equation}\label{d16}
{J}_{-}^{n}\ket{m_1=j_1,m_2=j_2}=\sum_{k=0}^{n}\binom{n}{k}{J_2}^{n-k}{J}_{-1}^{k}
\ket{m_1=j_1,m_2=j_2}.
\end{equation}
By integrating the expression obtained from the iterative application of the single angular momentum operator $J_{-}$, we derive the following formulation for Eq.\ref{d16}:
\begin{equation}\label{d17}
\begin{split}
    {J}_{-}^{n}\ket{m_1=j_1,m_2=j_2}
    &=\sum_{k=0}^{n}n!\hbar^{n}\sqrt{\binom{2j_1}{k}\binom{2j_2}{n-k}}\ket{m_1=j_1-k,m_2=j_2-n+k}.
\end{split}
\end{equation}
Hence, by substituting Eq.\eqref{d11} alongside Eq.\eqref{d17} into Eq.\eqref{d12}, the reconstruction of each state within the specified subspace is achieved through the subsequent equation:
\begin{equation}\label{d18}
    \ket{J=j_1+j_2,M=j_1+j_2-n}_J=\sum_{k=0}^{n}\sqrt{\frac{\binom{2j_1}{k}\binom{2j_2}{n-k}}{\binom{2j_1+2j_2}{n}}}\ket{m_1=j_1-k,m_2=j_2-n+k}.
\end{equation}

\subsection{Building the states with $J=j_1+j_2-n$}
We shall construct the general subspace of $J=j_1+j_2-m$. In this subspace, the state characterized by maximum angular momentum is represented by:
\begin{equation}\label{d19}
    \ket{J=j_1+j_2-m,M=j_1+j_2-m}_J=\sum_{l=0}^{m}\alpha_{l}\ket{m_1=j_1-l,m_2=j_2-m+l}.
\end{equation}
where $\alpha_{l}$ are undetermined constants. Given that ${J}_{+}\ket{j,m=j}=0$, it follows from Eq.\eqref{d19} that:
\begin{equation}\label{d20}
    \sum_{l=0}^{n}\alpha_{l}{J}_{+}\ket{m_1=j_1-l,m_2=j_2-m+l}=0.
\end{equation}
in which $J_{+}$ represents an operator articulated in the tensor product basis:
\begin{equation}\label{d21}
    {J_+}={J}_{+1}+{J}_{+2}.
\end{equation}
We shall examine the effect of ${J}_{+}$ when applied to an individual angular momentum state $\ket{j,m=-j}$:\begin{equation}\label{d22}
\begin{split}
    &{J}_{+}\ket{j,m=-j}=\hbar \sqrt{j(j-1)-(-j)(-j+1)}\ket{j,m=-j+1},\\
    &\vdots \\
    &{J}_{+}^{n}\ket{j,m=-j}=\hbar^{n}\sqrt{\prod_{k=0}^{n-1} j(j+1)-(-j+1)(-j+k+1)}\ket{j,m=-j+n}.
    \end{split}
\end{equation}
Upon simplifying the product in Eq.\eqref{d22}, we obtain:
\begin{equation}\label{d23}
    \prod_{k=0}^{n-1} j(j+1)-(-j+1)(-j+k+1)=(n!)^2 \binom{2j}{n}.
\end{equation}
Then the recurrent application of $J_{+}$ upon a state $\ket{j,m=-j}$ is expressed as follows:
\begin{equation}\label{d24}
    {J}_{+}^{n} \ket{j,m=-j}=\hbar^{n}n!\sqrt{\binom{2j}{n}}\ket{j,m=-j+n}.
\end{equation}
According to Eq.\eqref{d24}, the result of applying $J_{+}^{2j-l+1}$ to $\ket{j,m=-j}$ results in:
\begin{equation}\label{d25}
\begin{split}
    {J}_{+}^{2j-l+1}\ket{j,m=-j}
    &= \hbar^{2j-l+1}(2j-l+1)!\sqrt{\binom{2j}{2j-l+1}}\ket{j,m=j-l+1}.
\end{split}
\end{equation}
then:
\begin{equation}\label{d26}
    {J}_{+}\ket{j,m=j-l}=\hbar (2j-l+1) \sqrt{\frac{\binom{2j}{2j-l+1}}{\binom{2j}{2j-l}}}\ket{j,m=j-l+1}.
\end{equation}
The results from Eq.\eqref{d26} can be employed in Eq.\eqref{d20} to ascertain the application of $J_{+}=J_{+1}+J_{+2}$ to the states involved in the expression:
\begin{equation}\label{d27}
    {J}_{+1}\ket{j_1-l,j_2-m+l}=\hbar (2j_1-l+1)\sqrt{\frac{\binom{2j_1}{2j_1-l+1}}{\binom{2j_1}{2j_1-l}}}\ket{j_1-l+1,j_2-m+l}.
\end{equation}
\begin{equation}\label{d28}
    {J}_{+2} \ket{j_1-l,j_2-m+l}=\hbar(2j_2-m+l+1)\sqrt{\frac{\binom{2j_2}{2j_2-m+l+1}}{\binom{2j_2}{2j_2-m+l}}}\ket{j_1-l,j_2-m+l+1}.
\end{equation}
By substituting Eq.\eqref{d27} and Eq.\eqref{d28} into Eq.\eqref{d20}, the resulting expression is obtained as follows:
\begin{equation}\label{d29}
    \begin{split}
        &\sum_{l=0}^{n} \alpha_l (2j_1-l+1)\sqrt{\frac{\binom{2j_1}{2j_1-l+1}}{\binom{2j_1}{2j_1-l}}}\ket{j_1-l+1,j_2-m+l} \\
        &+\sum_{l=0}^{n}\alpha_l(2j_2-m+l+1) \sqrt{\frac{\binom{2j_2}{2j_2-m+l+1}}{\binom{2j_2}{2j_2-m+l}}}\ket{j_1-j,j_2-m+l+1} = 0.
    \end{split}
\end{equation}
By renaming the summation index in the second summation of Eq.\eqref{d29} to $k=l+1$, the following expression is derived:
\begin{equation}\label{d30}
    \begin{split}
        &\sum_{k=1}^{n+1}\alpha_{k-1}(2j_2-m+k)\sqrt{\frac{\binom{2j_2}{2j_2-m+k}}{\binom{2j_2}{2j_2-m+k+1}}}\ket{j_1-k+1,j_2-m+k}\\
        &\sum_{l=0}^{n} \alpha_l (2j_1-l+1)\sqrt{\frac{\binom{2j_1}{2j_1-l+1}}{\binom{2j_1}{2j_1-l}}}\ket{j_1-l+1,j_2-m+l}=0.
    \end{split}
\end{equation}
Considering that the ket states exhibit orthonormality, the sole requisite for fulfilling Eq.\eqref{d30} lies in ensuring that each coefficient corresponding to each ket within the summation is uniformly zero:
\begin{equation}\label{d31}
    (2j_1-l+1)\sqrt{\frac{\binom{2j_1}{2j_1-l+1}}{\binom{2j_1}{2j_1-l}}}\alpha_l+(2j_2-m+l)\sqrt{\frac{\binom{2j_2}{2j_2-m+l}}{\binom{2j_2}{2j_2-m+l+1}}}\alpha_{l-1}=0.
\end{equation}
This condition is satisfied for $\l=1,2,\hdots ,n$. Accordingly, we formulate a relationship that articulates $\alpha_l$ in terms of $\alpha_{l-1}$ as follows:
\begin{equation}\label{d32}
    \alpha_l=-\left(\frac{2j_2-m+l}{2j_1-l+1}\right)\sqrt{\frac{\binom{2j_2}{2j_2-m+l}\binom{2j_1}{2j_1-l}}{\binom{2j_2}{2j_2-m+l-1}\binom{2j_1}{2j_1-l+1}}}\alpha_{l-1}.
\end{equation}
By expanding the relation presented in Eq.\eqref{d32}, each coefficient $\alpha_l$ can be articulated in terms of $\alpha_0$:
\begin{equation}\label{d33}
    \alpha_{l} = (-1)^{l}\left[ \prod_{k=1}^{l} \left(\frac{2j_2-m+k}{2j_1-k+1} \right)\sqrt{\frac{\binom{2j_1}{2j_1-k}\binom{2j_2}{2j_2-m+k}}{\binom{2j_1}{2j_1-k+1}\binom{2j_2}{2j_2-m+k-1}}}\right] \alpha_0.
\end{equation}
The binomial coefficients as presented in Eq.\eqref{d33} can be reduced to:\begin{equation}\label{d34}
    \alpha_l=(-1)^{l} \left[\prod_{k=1}^{l}\sqrt{\frac{(2j_2-m+k)(m-k+1)}{k(2j_1-k+1)}} \right] \alpha_0.
\end{equation}
The value of $\alpha_0$ is determined by imposing the normalization condition on the state:
\begin{equation}\label{d35}
    \alpha_0=\frac{1}{\sqrt{\sum_{l=0}^{m} \prod_{k=1}^{l} \frac{(2j_2-m+k)(m-k+1)}{k(2j_1-k+1)}}}.
\end{equation}
where we take the convention:
\begin{equation} \label{d36}
    \prod_{k=1}^{l} \frac{(2j_2-m+k)(m-k+1)}{k(2j_1-k+1)} \Big{|}_{l=0} = 1.
\end{equation}
By inserting Eq.\eqref{d35} into Eq.\eqref{d18}, the general expression for the state of maximum angular momentum within the subspace $J=j_1+j_2-m$ is derived:
\begin{equation}\label{d37}
    \ket{j_1+j_2-m,j_1+j_2-m}_{J} = \sum_{l=0}^{m} \frac{(-1)^{l}\prod_{k=1}^{l} \sqrt{\frac{(2j_2-m+k)(m-k+1)}{k(2j_1-k+1)}}}{\sqrt{\sum_{i=0}^{m} \prod_{q=1}^{i} \frac{(2j_2-m+q)(m-q+1)}{q(2j_1-q+1)}}} \ket{j_1-l,j_2-m+l}.
\end{equation}
By explicitly working the product factor appearing in Eq.\eqref{d37} we can simplify the expression into the following:
\begin{equation}\label{d371}
    \prod_{k=1}^{l} \frac{(2j_2-m+k)(m-k+1)}{k(2j_1-k+1)}= \frac{\binom{2j_2-m+k}{k}\binom{m}{k}}{\binom{2j_1}{k}}
\end{equation}
By applying operator ${J}_{-}$ to Eq.\eqref{d37} and substituting Eq.\eqref{d371} for the product factor, the subspace for $J=j_1+j_2-m$ can be reconstructed:
\begin{equation}\label{d38}
\begin{split}
    &{J}_{-}^{s}\ket{j_1+j_2-m,j_1+j_2-m}_J = \hbar^s s! \sqrt{\binom{2j_1+2j_2-2n}{s}}\ket{j_1+j_2-m,j_1+j_2-m-s}_J,\\
    & {J}_{-}^s\ket{m_1=j_1-l,m_2=j_2-m+l}=\sum_{p=0}^{s} \alpha(p)\ket{j_1-l-p,j_2-m+l-s+p}.
    \end{split}
\end{equation}
in which the coefficient denoted by $\alpha(p)$ is specified as follows:\begin{equation}
   \alpha(p)= \hbar^s \frac{(l+p)!(m-l+s-p)!}{l!(m-l)!}\binom{s}{p}\sqrt{\frac{\binom{2j_1}{l+p}\binom{2j_2}{m-l+s-p}}{\binom{2j_1}{l}\binom{2j_2}{m-l}}}.
\end{equation}
Based on Eq.\eqref{d38} and following extensive simplification of the derived factors, we obtain the subsequent expression for the state within the subspace:
\begin{equation}\label{d39}
    \ket{j_1+j_2-m,j_1+j_2-m-s}_{J} = \sum_{l=0}^{m}\sum_{p=0}^{s} \beta(l,p) \ket{j_1-l-p,j_2-m+l-s+p}.
\end{equation}
in which the specification for $\beta(l,p)$ is given by:\begin{equation}
    \beta(l,p)=(-1)^{l}\frac{\sqrt{\frac{\binom{2j_1-l}{p}\binom{2j_2-m+l}{s-p}\binom{2j_2-m+l}{l}\binom{l+p}{p}\binom{m-l+s-p}{s-p}\binom{m}{l}}{\binom{2j_1}{l}\binom{2j_1+2j_2-2m}{s}}}}{\sqrt{\sum_{i=0}^{m} \frac{\binom{2j_2-m+i}{i}\binom{m}{i}}{\binom{2j_1}{i}}}}.
\end{equation}
Equation\eqref{d39} represents the general expression for the states within the given subspace, therefore it reconstructs the subspace.
\subsection{Clebsch-Gordan Coefficients}
Recasting the eigenvalues of the general state in Eq.\eqref{d39} into a generalized format $J=j_1+j_2-m,M=j_1+j_2-m-s$, we can rewrite Eq.\eqref{d39} as follows:
\begin{equation}\label{d40}
    \ket{J,M}_J=\sum_{l=0}^{j_1+j_2-J}\sum_{m_1=j_1-l-J+M}^{j_1-l}\beta(l,m_1)\ket{m_1,M-m_1}.
\end{equation}
where:
\begin{equation}
\beta(l,m_1)=(-1)^l \sqrt{\frac{\binom{2j_1-l}{j_1-m_1-l}\binom{j_2-j_1+J+l}{J-M+m_1-j_1+l}\binom{j_2-j_1+J+l}{l}\binom{j_1-m_1}{l}\binom{j_2-M+m_1}{J-M+m_1-j_1+l}\binom{j_1+j_2-J}{l}}{\binom{2j_1}{l}\binom{2J}{J-M} \sum_{i=0}^{j_1+j_2-J}\frac{\binom{j_2-j_1+J+i}{i}\binom{j_1+j_2-J}{i}}{\binom{2j_1}{i}}}}.
\end{equation}
The projection onto state $\bra{m_1',M-m_1'}$ yields the general Clebsch-Gordan coefficient for this state:
\begin{equation}\label{d41}
    C_{m_1,M-m_1}=\sum_{l=K}^{N}(-1)^l \sqrt{\frac{\binom{2j_1-l}{j_1-m_1-l}\binom{j_2-j_1+J+l}{J-M+m_1-j_1+l}\binom{j_2-j_1+J+l}{l}\binom{j_1-m_1}{l}\binom{j_2-M+m_1}{J-M+m_1-j_1+l}\binom{j_1+j_2-J}{l}}{\binom{2j_1}{l}\binom{2J}{J-M} \sum_{i=0}^{j_1+j_2-J}\frac{\binom{j_2-j_1+J+i}{i}\binom{j_1+j_2-J}{i}}{\binom{2j_1}{i}}}}.
\end{equation}
where:
\begin{equation}
    K=\max(0,j_1-J+m_2), \quad N=\min(j_1+j_2-J,j_1-m_1).
\end{equation}
Subsequently, the general expression for the Clebsch-Gordan coefficient is articulated by the following formula:
\begin{equation}\label{d42}
C_{m_1,m_2}=\delta_{m_1+m_2,M}\sum_{l=K}^{N}(-1)^l \sqrt{\frac{\binom{2j_1-l}{j_1-m_1-l}\binom{j_2-j_1+J+l}{J-M+m_1-j_1+l}\binom{j_2-j_1+J+l}{l}\binom{j_1-m_1}{l}\binom{j_2-M+m_1}{J-M+m_1-j_1+l}\binom{j_1+j_2-J}{l}}{\binom{2j_1}{l}\binom{2J}{J-M} \sum_{i=0}^{j_1+j_2-J}\frac{\binom{j_2-j_1+J+i}{i}\binom{j_1+j_2-J}{i}}{\binom{2j_1}{i}}}}.
\end{equation}

Equation\eqref{d42} shows that the only non-zero Clebsch-Gordan coefficients are those for which $M=m_1+m_2$ which is in agreement with the existing formulas obtained by Racah\cite{Racah1942}, Wigner\cite{wigner2012group} and Schwinger\cite{schwinger1960angular}. By construction, Eq.\eqref{d42} has the same symmetry properties of Clebsch-Gordan coefficients. However, it has a non-trivial different form to those previously obtained in the literature. Giving a mathematical rigorous proof of the equivalence of the obtained formula in Eq.\eqref{d42} to that of Racah and Wigner is very challenging and out of scope of this work. The initial obtained expressions from Racah\cite{Racah1942} and Wigner\cite{wigner2012group} didn't have the same form, but through extensive algebraic manipulations was possible to show the mathematical equivalence. Thus let us convince ourselves of the agreement by using it in example cases and compare its results versus already established results in literature. To do so, we provide a Python implementation of Eq. \eqref{d42} which allows us to calculate the Clebsch-Gordan coefficients by using this formula and compare them against Clebsch-Gordan coefficients computed through the SymPy package\cite{sympy2025cg}.

\newpage
\subsubsection{$j_1=2,j_2=1$}
In the following table are listed the obtained Clebsch-Gordan Coefficients with the Python implementation of Eq.\eqref{d42} and the SymPy implementation for the addition of $j_1=2,j_2=1$.
\begin{table}[h!]
\centering
\caption{Comparison of Clebsch–Gordan Coefficients $j_1=2,j_2=1$: Alternative Formula vs SymPy}
\scalebox{0.6}{
\begin{ruledtabular}
\begin{tabular}{rrrrrr}
\textbf{$J$} & \textbf{M} & \textbf{$m_1$} & \textbf{$m_2$} & \textbf{Alternative Formula} & \textbf{SymPy} \\
1 & -1 & -2 & 1  & 0.77460 & 0.77460 \\
1 & -1 & -1 & 0  & -0.54772 & -0.54772 \\
1 & -1 & 0  & -1 & 0.31623 & 0.31623 \\
1 & 0  & -1 & 1  & 0.54772 & 0.54772 \\
1 & 0  & 0  & 0  & -0.63246 & -0.63246 \\
1 & 0  & 1  & -1 & 0.54772 & 0.54772 \\
1 & 1  & 0  & 1  & 0.31623 & 0.31623 \\
1 & 1  & 1  & 0  & -0.54772 & -0.54772 \\
1 & 1  & 2  & -1 & 0.77460 & 0.77460 \\
2 & -2 & -2 & 0  & -0.81650 & -0.81650 \\
2 & -2 & -1 & -1 & 0.57735 & 0.57735 \\
2 & -1 & -2 & 1  & -0.57735 & -0.57735 \\
2 & -1 & -1 & 0  & -0.40825 & -0.40825 \\
2 & -1 & 0  & -1 & 0.70711 & 0.70711 \\
2 & 0  & -1 & 1  & -0.70711 & -0.70711 \\
2 & 0  & 1  & -1 & 0.70711 & 0.70711 \\
2 & 1  & 0  & 1  & -0.70711 & -0.70711 \\
2 & 1  & 1  & 0  & 0.40825 & 0.40825 \\
2 & 1  & 2  & -1 & 0.57735 & 0.57735 \\
2 & 2  & 1  & 1  & -0.57735 & -0.57735 \\
2 & 2  & 2  & 0  & 0.81650 & 0.81650 \\
3 & -3 & -2 & -1 & 1.00000 & 1.00000 \\
3 & -2 & -2 & 0  & 0.57735 & 0.57735 \\
3 & -2 & -1 & -1 & 0.81650 & 0.81650 \\
3 & -1 & -2 & 1  & 0.25820 & 0.25820 \\
3 & -1 & -1 & 0  & 0.73030 & 0.73030 \\
3 & -1 & 0  & -1 & 0.63246 & 0.63246 \\
3 & 0  & -1 & 1  & 0.44721 & 0.44721 \\
3 & 0  & 0  & 0  & 0.77460 & 0.77460 \\
3 & 0  & 1  & -1 & 0.44721 & 0.44721 \\
3 & 1  & 0  & 1  & 0.63246 & 0.63246 \\
3 & 1  & 1  & 0  & 0.73030 & 0.73030 \\
3 & 1  & 2  & -1 & 0.25820 & 0.25820 \\
3 & 2  & 1  & 1  & 0.81650 & 0.81650 \\
3 & 2  & 2  & 0  & 0.57735 & 0.57735 \\
3 & 3  & 2  & 1  & 1.00000 & 1.00000 \\
\end{tabular}
\end{ruledtabular}}
\end{table}\\
As can be seen the obtained coefficients are exactly the same.
\subsubsection{$j_1=1,j_2=1$}
In the following table are listed the obtained Clebsch-Gordan Coefficients witht the Python implementation of Eq.\eqref{d42} and the SymPy implementation for the addition of $j_1=1,j_2=1$.
\begin{table}[H]
\centering
\caption{Comparison of Clebsch–Gordan Coefficients $j_1=1,j_2=1$: Alternative Formula vs SymPy}
\scalebox{0.70}{
\begin{ruledtabular}
\begin{tabular}{rrrrrr}
\textbf{$J$} & \textbf{M} & \textbf{$m_1$} & \textbf{$m_2$} & \textbf{Alternative Formula} & \textbf{SymPy} \\
0 & 0  & -1 & 1  & 0.57735 & 0.57735 \\
0 & 0  & 0  & 0  & -0.57735 & -0.57735 \\
0 & 0  & 1  & -1 & 0.57735 & 0.57735 \\
1 & -1 & -1 & 0  & -0.70711 & -0.70711 \\
1 & -1 & 0  & -1 & 0.70711 & 0.70711 \\
1 & 0  & -1 & 1  & -0.70711 & -0.70711 \\
1 & 0  & 1  & -1 & 0.70711 & 0.70711 \\
1 & 1  & 0  & 1  & -0.70711 & -0.70711 \\
1 & 1  & 1  & 0  & 0.70711 & 0.70711 \\
2 & -2 & -1 & -1 & 1.00000 & 1.00000 \\
2 & -1 & -1 & 0  & 0.70711 & 0.70711 \\
2 & -1 & 0  & -1 & 0.70711 & 0.70711 \\
2 & 0  & -1 & 1  & 0.40825 & 0.40825 \\
2 & 0  & 0  & 0  & 0.81650 & 0.81650 \\
2 & 0  & 1  & -1 & 0.40825 & 0.40825 \\
2 & 1  & 0  & 1  & 0.70711 & 0.70711 \\
2 & 1  & 1  & 0  & 0.70711 & 0.70711 \\
2 & 2  & 1  & 1  & 1.00000 & 1.00000 \\
\end{tabular}
\end{ruledtabular}}
\end{table}
As can be seen the obtained coefficients are exactly the same.
Although not reproduced here, through the Python implemantion of Eq.\eqref{d42} we tested the results of Clebsch-Gordan coefficients up to $j_1=20,j_2=20$ and we obtained exactly the same results as the SymPy implementation. This supports the validity and agreement of the obtained formula with standard literature.
\section{conclusions}\label{s3}
The general formula for the Clebsch-Gordan coefficients presented in Eq.\eqref{d42} serves as an alternative method for calculating these coefficients, alongside existing formulas in the literature. The novelty of rebuilding non-maximal angular momentum subspaces through raising operators allowed us to obtain an alternative formula for Clebsch-Gordan coefficients. Despite its complexity and length—rendering it more challenging than Racah's formula it offers a significant advantage as it has been derived entirely through elementary raising and lowering operator methods of Quantum Mechanics, without necessitating additional mathematical expertise. This derivation enhances its accessibility to a wider audience of physicists.
\\Although agreement with standard literature was observed through code implementation, it is imperative to give the rigorous proofs establishing mathematical equivalence among previously identified formulas in the literature as a subsequent endeavor.

\appendix*   
\section{Python Code for Clebsch–Gordan Coefficients}

The Python script below gives an implementation of the derived alternative formula for Clebsch-Gordan coefficients given by Eq.\eqref{d42} and it compares it with SymPy implementation given in the Sympy package\cite{sympy2025cg} to compute Clebsch–Gordan coefficients. The output is saved as an Excel file.

\begin{lstlisting}[language=Python, breaklines=true,frame=single]
#Import nedeed libraries
# SymPy library is used to calculate standard CG Coefficients.
import numpy as np
import math
from scipy.special import binom
from sympy.physics.quantum.cg import CG
import pandas as pd

# Function to compute CG Coefficients according to derived alternative formula given in Eq. (69) of the article.
# J: total angular momentum
# M: z component of total angular momentum
# j_1: angular momentum of subsystem 1
# j_2: angular momentum of subsystem 2
# m_1: z component of angular momentum of subsystem 1
# m_2: z component of angular momentum of subsystem 2
def clebsch_gordan_coef(J,M,j_1,j_2,m_1,m_2):
    K=max(0,j_1-J+M-m_1)
    N=min(j_1+j_2-J,j_1-m_1)
    if M==m_1+m_2:
        coef=0
        for l in range(K,N+1):
            coef=coef+((-1)**(l)*np.sqrt(binom(2*j_1-l,j_1-m_1-l)*binom(j_2-j_1+J+l,J-M+m_1-j_1+l)*binom(j_2-j_1+J+l,l)*binom(j_1-m_1,l)*binom(j_2-M+m_1,J-M+m_1-j_1+l)*binom(j_1+j_2-J,l)/(binom(2*j_1,l)*binom(2*J,J-M))))
        normalization=0
        for i in range(0,j_1+j_2-J+1):
            normalization = normalization+binom(j_2-j_1+J+i,i)*binom(j_1+j_2-J,i)/(binom(2*j_1,i))
        coef=coef/np.sqrt(normalization)
    else:
        coef=0
    return coef

# For loop to calculate the Clebsch Gordan Coefficients via alternative formula found and SymPy implementation.
# This loop calculates the coefficients up to a total angular J=j_1+j_2.

list_of_results=[]
j_1=4
j_2=3
j_max=max(j_1,j_2)
j_min = min(j_1,j_2)
for J in range(j_max-j_min,j_max+j_min+1):
    for M in range(-J,J+1):
        for m_1 in range(-j_1,j_1+1):
            for m_2 in range(-j_2,j_2+1):
                coef = clebsch_gordan_coef(J,M,j_1,j_2,m_1,m_2)
                coef1 = CG(j_1,m_1,j_2,m_2,J,M).doit().evalf()

                if np.isnan(coef):
                    coef=0

                if coef!=0:
                    if abs(coef1-coef)<1e-4:
                        y="equal"
                    else:
                        y="different"
                    print("J=",J,"M=",M,"m_1=",m_1,"m_2=",m_2,"CG Coefficient Alternative Derived Formula=",f"{coef:.5f}",",CG Coefficient using SymPy=",f"{coef1:.5f}")
                    list_of_results.append([J,M,m_1,m_2,f"{coef:.5f}",f"{coef1:.5f}"])
    
# Store the obtained coefficients in a pandas dataframe
dataframe = pd.DataFrame(list_of_results)
dataframe.columns = ["J","M","m1","m2","CG Coefficient Alternative Formula","CG Coefficient SymPy"]
dataframe.to_excel("CG Table.xlsx",index=False)
\end{lstlisting}

\bibliography{biblio}
\bibliographystyle{plain}

\end{document}